\documentclass[]{aa}  
\usepackage{graphicx}
\usepackage{xcolor}
\usepackage{txfonts}
\usepackage{natbib}
\usepackage{placeins}
\usepackage[modulo,switch]{lineno}
\bibpunct{(}{)}{;}{a}{}{,} 
\usepackage{amsfonts} 
\usepackage{amsmath}
\usepackage{amssymb}
\usepackage{textcomp, gensymb} 
\usepackage{wasysym}  
\usepackage{mathrsfs}
\usepackage{siunitx} 
\usepackage{appendix}
\usepackage{hyperref} 

\def\cd  {{$\mbox{c~d}^{-1}$}}

\def\top{$T_{\rm eff}$}
\def\gop{$\log g$}

\def\msun {{\mathrm{M}_\odot}}

\def\simless{\mathbin{\lower 3pt\hbox
     {$\rlap{\raise 5pt\hbox{$\char'074$}}\mathchar"7218$}}}   
\def\simmore{\mathbin{\lower 3pt\hbox
     {$\rlap{\raise 5pt\hbox{$\char'076$}}\mathchar"7218$}}}   

\def\cd  {{$\mbox{d}^{-1}$}}

\begin{document}

   \title{Fast optical variability in supergiant X-ray binaries \\
   }

   \author{
D. Di Filippantonio\inst{1},
P. Reig\inst{2,3}, \and
J. Fabregat\inst{4}
}

   \institute{Universidad Internacional de Valencia, Calle Pintor Sorolla, 21, 46002, Valencia, España \\
\email{denis.df@gmail.com}
\and
Institute of Astrophysics, Foundation for Research and Technology, 71110 Heraklion, Crete, Greece\\
\email{pau@physics.uoc.gr}
\and
University of Crete, Physics Department \& Institute of
Theoretical \& Computational Physics, 70013 Heraklion, Crete, Greece
\and
Observatorio Astron\'omico de la Universidad de
Valencia, Calle Catedr\'atico Jos\'e Beltr\'an 2, 46980 Paterna, Valencia, Spain \\
\email{juan.fabregat@uv.es}
}

   \date{Received ; accepted }

 
\abstract 
{Recent studies of massive stars using high-precision space photometry have
revealed that they commonly exhibit stochastic low-frequency variability.
}
{The main goal of this work is to investigate the fast photometric variability of
the optical counterparts to supergiant X-ray binaries and to compare the general
patterns of this variability with that observed in the Galactic population of other early-type 
stars. 
}
{ We selected a sample of 14 high-mass X-ray binaries with supergiant companions observed by
the Transiting Exoplanet Survey Satellite (TESS). We also studied 4
Be/X-ray binaries with persistent X-ray emission for comparison. The TESS light
curves were created from the full-frame images using the {\tt Lightkurve} 
package.  The light curves were background subtracted and corrected for
scattered light and instrumental effects. Standard Fourier analysis was used to
obtain the periodograms. We used a phenomenological model to fit the amplitude
spectra and derive the best-fit parameters. 
} 
{ All sources exhibit fast aperiodic light variations. The shape of the
periodogram is well described by a red noise component at intermediate
frequencies ($\sim 1-10$ \cd). At lower frequencies, the noise level flattens, while at higher frequencies the periodogram is dominated by white noise. We find
that the patterns of variability of the massive companions in supergiant X-ray
binaries agree with those of single early-type  evolved stars in terms of the
general shape of the periodograms.  However, they exhibit higher amplitude at
low frequencies and lower characteristic frequencies than those of Be/X-ray binaries.
Unlike Be/X-ray binaries, supergiant X-ray binaries exhibit a total lack of
coherent signals at high frequencies. Most sources have been analyzed over
multiple TESS sectors, spanning a duration of 4 years. We do not find any
significant variation over time in the low-frequency variability
characteristics. 
}
{This study reveals that stochastic low-frequency variability is a very common,
if not ubiquitous, feature intrinsic to supergiant optical companions in X-ray
binaries. The phenomenology of this variability is similar to that of single
early-type supergiant stars. }


   \maketitle
   
%

\section{Introduction}

High-mass X-ray binaries (HMXBs) are divided into supergiant X-ray binaries
(SGXBs) and Be/X-ray binaries (BeXBs) according to the luminosity of the optical
counterpart. In a SGXB \citep{martinez17,kretschmar21}, a neutron star orbits around an evolved star (luminosity
class I or II),  while in BeXBs \citep{reig11,paul11} the optical companion is a 
main sequence star (luminosity class III--V). In Be stars, episodic ejection of
mass from their atmospheres results in the formation of a flattened disk around
their equator \citep{porter03,rivinius13a}. 
Rotation and pulsations are characteristics of the fast optical variability in
BeXBs and manifest as multifrequency photometric variability
\citep{baade16,rivinius16,semaan18,balona20,balona21} and as spectroscopic
variability, for example in the form of line-profile variations
\citep{baade84,rivinius98,rivinius03,balona03,zima06} with typical periods in
the range of 0.1 to 2 days. In SGXBs, the supergiant star produces powerful stellar winds that supply the material accreted onto the compact object. Most known SGXB systems transfer mass from the donor to the neutron star via this mechanism. However, in a handful of systems, the mass transfer occurs most likely via Roche lobe overflow (RLOF), which leads to a higher mass flow. In such cases, an accretion disk is formed around the neutron star and the X-ray emission is increased \citep{corbet86,kretschmar19}.

A subclass of SGXBs are the supergiant fast X-ray transients (SFXT), which are
similar to the other SGXBs in the sense that they contain an OB supergiant as
the mass donor and that the mass-transfer mechanism is through stellar wind. The
difference lies in the fact that, while other SGXB systems show a relatively
constant emission in the X-ray range, the SFXTs present transient X-ray emission
with fast outbursts, corresponding to a variable amount of material being
transferred from the supergiant to the compact object \citep{negueruela06,
kretschmar19}.

The advent of space-based high-accuracy photometry missions, such as \textit{Kepler} \citep{koch10} and TESS \citep{ricker15}, has enabled detailed studies of the short-term variation of stellar luminosity, which, through asteroseismology, allows us to probe the internal structure of the star. In particular, the detection and analysis of the coherent pulsation modes and stochastic low-frequency (SLF) variability of massive stars have only recently  been made possible through this new set of long-duration, high-accuracy photometry measurements \citep{aerts17,tkachenko14,bowman19,dorn-wallenstein19,bowman20,dorn-wallenstein20}. 

In \citet{reig22}, we studied the fast optical variability of BeXBs. The motivation behind that study was to investigate if the same phenomenology  is observed in the short-term variability of Be stars regardless of whether they belong to an X-ray binary or are single systems. In other words, our aim was to find out whether or not the neutron star affects the rapid optical variability properties. We found that BeXBs and classical Be stars are indistinguishable in terms of their pulsational characteristics. 

SGXBs have shorter orbital periods than BeXBs, and therefore the neutron star passes closer to the optical companion \citep{kim23}. In a similar approach to that applied to BeXBs, in the present work we intend to compare the rapid optical variability of massive companions in X-ray binary systems with that of single supergiant stars. Likewise, we compare the variability properties of SGXBs with persistent (in X-rays) BeXBs. The X-ray luminosity of persistent BeXBs is two or three orders of magnitude lower than of transient BeXBs during outbursts, indicating that they accrete from the stellar wind of the companion and/or from the outer parts of a highly debilitated disk and are not expected to show a strong neutron-star--disk interaction. Persistent BeXBs exhibit more stable mass loss, and display optical light curves similar to those of SGXBs in terms of their variability properties \citep{reig99}.

Unlike Be stars, whose amplitude spectra are dominated by coherent pulsation modes attributed to non-radial pulsations caused by changes in the opacity and opacity gradient in different stellar layers \citep[][see also \citealt{aerts21} for a review]{pamyatnykh99}, SLF variability in massive evolved stars is caused by some kind of turbulence or instability within the star, such as waves excited from near the convective core and/or near the surface \citep[see][for a recent review]{bowman23}.
This work focuses on the analysis of SLF variability in SGXB systems, and their comparison to BeXBs and isolated supergiants. 

\begin{table*}
\label{list}
\caption{Selected list of sources}      
\centering          
\begin{tabular}{l@{~~}l@{~~}c@{~~}c@{~~}c@{~~}c@{~~}c@{~~}c@{~~}l@{~~}}
\hline\hline
Object          &        Type   &        Sp.Typ.        &        G (mag)         &       $P_{orb}$ [d]   &       $P_{spin}$ [s]  &       Mass ($\msun$)  &        TESS ID         &        TESS sectors           \\
\hline
IGR J00370+6122         &       SGXB    &        BN0.7Ib        &        9.5             &       15.665  &       674     &       22      &        284207391       &        17,18,24,58          \\
2S 0114+650     &       SGXB    &        B1Iae          &        10.5   &       11.598  &       10008   &       16      &        54469882        &        18,24,25,52,58       \\
Vela X-1                &       SGXB    &        B0.5Iae        &        6.7             &       8.963   &       283     &       26      &        191450569       &        08,09,35,62            \\
Cen X-3                 &       SGXB    &        O6-7II-III     &        12.9    &       2.033   &       4.802   &       20.2    &        468240308         &        10,11,37,64           \\
1E 1145.1-6141  &       SGXB    &        B2Iae          &        12.3   &       14.365  &       298     &       14      &        324122933       &        10,11,37,38,64       \\
4U 1538-522     &       SGXB    &        B0.2Ia         &        13.2   &       3.728   &       526.4   &       20      &        190415214       &        12,39,65              \\
4U 1700-377     &       SGXB    &        O6Ia           &        6.4             &       3.412   &       --      &       46      &        347468930         &        12,39,66              \\
Cyg X-1                 &       SGXB    &        O9.7Iab        &        8.5             &       5.599   &       --      &       40.6    &        102604645       &        14,54,55,74,75       \\
IGR J08408-4503         &       SFXT    &        O8.5Ib-II      &        7.5             &       9.54    &       --      &       33      &        141856104       &        08,09,35,61,62       \\
IGR J11215-5952         &       SFXT    &        B1Ia           &        9.8             &       164.6   &       186.8   &       --      &        451004619       &        10,11,37,64           \\
IGR J16465-4507         &       SFXT    &        B0.5-1Ib       &        13.5    &       30.24   &       228     &       27.8    &        236510713         &        12,39,66            \\
XTE J1739-302   &       SFXT    &        O8Iab          &        12.6   &       51.47   &       --         &       33.7    &        106579192      &        39                   \\
RX J0146.9+6121 &       BeXB    &        B1IIIe         &        11.2   &       330     &       1407.4  &       9.6     &        399437313       &        58                     \\
X Per                   &       BeXB    &        B1Ve           &        6.3             &       250.3   &       837.7   &       15.5    &        94471007        &        18,43,44,70,71       \\
4U 1036-56              &       BeXB    &        B0III-Ve       &        11.2    &       60.9    &       860     &       17.5    &        458199429         &        10,36,37,63,64       \\
4U 2206+543     &       BeXB    &        O9.5Ve         &        9.7             &       9.558   &       392     &       18      &        328546890         &        16,17,56,57            \\
XTE J0421+560   &       sgB[e]  &        B0/2I[e]       &        10.8   &       19.41   &       --         &       --      &        418090700      &        19,59,73             \\
4U 1954+319     &       sg(M)   &        MI             &        8.4             &       1296.6  &       18612   &       9       &        87456211         &        14,41,54,55,74,75       \\
\hline\hline
\end{tabular}
\tablefoot{Periods and mass from \citet{fortin23}}  
\end{table*}
 
\section{Observations}
\label{obs}

We used data assembled by the Transiting Exoplanet Survey Satellite (TESS). The primary objective of TESS is to discover new exoplanets using the transit method and the instrument is optimized for the detection of super-Earth-sized planets around nearby bright stars \citep{ricker15}. However, owing to its observational strategy, TESS covers the entire sky in periods of two years. This makes TESS an excellent mission for monitoring the variability of many different types of objects. In particular, TESS has opened up the sky for massive studies of  stellar variability (e.g., \citealt{burssens20}), where previous missions focused generally on low-mass stars.

The TESS instrument is comprised of four cameras with a wide field of view of 24$^{\circ}$ by 24$^{\circ}$ , which means a strip of 24$^{\circ}$ by 96$^{\circ}$ is observed at any given time. The cameras and CCD detectors are optimized for the red band, and are sensitive in the 600 to 1000 nm range, as the primary mission objective is to survey stars with spectral types from F5 to M5. Each camera is composed of 4 CCDs of 2048x2048 pixels, corresponding to a relatively low spatial resolution of 21 arcsec per pixel.

The CCDs are read out at two-second intervals and are then stacked on board to produce lower-frequency data to be downlinked to Earth. Small windows around a predefined list of targets of interest (TOIs) are stacked over a duration of 2 minutes, while the complete CCD readout in the form of full-frame images (FFIs) are stacked over a longer duration. During the primary mission (July 2018 to July 2020, sectors 1 to 26), the FFIs were sampled with a cadence of 30 minutes; during the first mission extension (July 2020 to August 2022, sectors 27 to 55), this cadence was 10 minutes; and for the second mission extension (since September 2022, sector 56), this cadence was 200 seconds.
The spacecraft observes the same direction of the sky for two orbits, or 27.4 days, with this being referred to as one sector, before moving to a new observation region.


\section{Methodology}

This section describes the overall methodology and the various steps followed from obtaining the TESS science data to computing the best-fit model parameters.

\subsection{Selection of sources}

To compile the list of HMXBs to be analyzed, we referred to consolidated catalogs of HMXBs \citep{kim23,fortin23,neumann23}. In particular, \citet{fortin23} produced a catalog of galactic HMXBs with confirmed optical counterparts, which also includes a classification by HMXB type, simplifying the selection
of objects. As the focus of the present project is on SGXBs, the list of objects was initially filtered to only include objects whose optical companion is a supergiant star. This selection resulted in a total of 54 objects.

The next step was to check if these sources had a clear optical counterpart bright enough to extract a light curve using TESS FFI data, and far enough from any bright source so as to be separated in the TESS data, which have a limited spatial resolution (21 arcsec per pixel). Of the 54 preselected sources, 12 were not observed by TESS and another 20 were too faint to extract a useful light curve. Finally, 8 objects were eliminated due to their proximity ($<$ 2 pixels) to another object as bright or brighter ($>$ 1 mag), which prevented us from extracting a clean light curve. 

A total of 14 SGXBs remained in the list of objects to be analyzed, which included two peculiar systems with supergiant companions, even though they are not typical SGXBs. These are XTE J0421+560 with a B[e] supergiant and 4U 1954+319 with a later-type M supergiant star.
Four persistent BeXB sources were added to the list in order to compare them with the results obtained for the SGXBs. 

The final list of objects is shown in Table~\ref{list}, where the sources are sorted first by type and then by right ascension (RA). For each object, the list of sectors that contain valid data is provided. This amounts to a total of 70 sectors for all sources combined.



\subsection{Data processing}




We created our own light curves from the FFIs using the {\tt LIGHTKURVE}
\citep{lightkurve} and {\tt TESScut} \citep{brasseur19} packages to download a
target pixel file (TPF) with an $N\times N$ pixel image centered on the target
for every available {\it TESS} sector.  The size of the cutout is important for
the processing of the light curve in the subsequent steps. In particular, it
should be large enough to include enough background pixels to correct the raw
light curve for scattered light  ---in addition to the aperture mask discussed
in the following section. However, it should also be kept in mind that larger
cut-outs will lead to excessive amounts of data to download and process without
any advantage to the final light curve. After several iterations with various
objects, a cutout size of 15 by 15 pixels (5.25 $\times$ 5.25 arcmin$^2$) was
found to be adequate and was used for all sources.

The raw light curve was computed from the TPF by summing the flux over a defined
aperture mask for each cadence, and obtaining a total flux over time for the
whole duration of the sector. For each object, it is therefore necessary to
define an aperture mask that best covers the target object without including
bright contaminating sources.

A source mask was defined from pixels in the TPF, which show a median flux that
is brighter than the threshold multiplied by the standard deviation above the
overall median. Generally, we used a threshold of S/N=15 for brighter sources
($G<10$) but this was decreased to $S/N=5$ for fainter sources. Custom-made
masks were defined for  1E 1145.1-6141, Cen X-3, IGR J16465-4507, XTE J1739-302,
and Cyg X-1 to avoid including light from nearby stars. The background mask was
defined avoiding field sources (i.e. pixels with fluxes below the null
threshold).

The light curves were also corrected for scattered light using the regression
corrector
method\footnote{https://heasarc.gsfc.nasa.gov/docs/tess/NoiseRemovalv2.html}.
Finally, because the goal of this project is not to analyse long-term variations
of the sources but to study the SLF variability, long-term trends were removed
from the light curves when present. None of the SGXB, SFXT, or wind-fed BeXB
sources (4U 2206+543 and 4U 1026-56) showed any noticeable long-term variability on the
sectors observed. In contrast, the BeXB sources X Per and RX J0146.9+6121 show
long-term variability in all the analyzed sectors. Long-term variability is a
common behavior of Be stars, with slow quasi-periodic variation over periods of
month to years \citep{reig22}.

\subsection{Extraction of significant frequencies and pre-whitening}
\label{whitening}

Significant coherent signals are not expected in luminous SGXBs (type Ia), although they may be present in less luminous systems (type Ib or II).  \citet{bowman19} assembled a sample of 167 hot massive stars (91 supergiants and 76 stars with luminosity class III-V or unknown luminosity) with the purpose of investigating the incidence of coherent modes and stochastic variability.
Only two supergiant systems showed $\beta$ Cephei-like pulsations, while all (except one) type Ia supergiants (29 systems) exhibited a frequency spectrum consistent with SLF variability without any other kind of coherent modulation. Only one system presented a modulation that could be attributed to rotation by a clumpy aspherical wind. 
In contrast, almost all BeXBs display multi-periodic light modulations attributed to non-radial pulsations \citep{reig22}. 

To remove the periodic component, we performed the following steps (a process known as pre-whitening; see e.g., \citealt[][and references therein]{bowman21}):

\begin{enumerate}
\item Compute the periodogram.
\item Extract the frequency of the highest peak $f_{max}$.
\item Fit the time series with a function of the type 
$$A \sin(2\pi f_{max} t + \phi).$$
\item Subtract the fitted curve from the time series to obtain the residuals.
\item Compute the periodogram of the residuals.
\item Compute the signal-to-noise ratio of the peak, where the signal is the amplitude $A$ of the peak and the noise is the mean level of the surrounding continuum.
\item If the peak is significant (S/N > 4), we annotate its frequency and amplitude and proceed again to step 1.
\end{enumerate}

This iterative process allows us to detect all significant pulsations and obtain a pre-whitened light curve, which should only contain random noise components, or at least a combination of non-significant frequencies. We note that some authors \citep[see e.g.,][]{bowman21} have warned against the use of four times the noise level as the optimum criterion to extract significant frequencies, arguing that this is too low, and includes some fraction of peaks whose origin is simply white noise. Nevertheless, as we compare our work with previous works that used the same criterion, we opted to use S/N>4.

Another critical decision is the width of the frequency interval around the peak to compute the noise level. An overly broad interval would include mostly the white noise component at higher frequency, and all the apparent peaks at low frequency would appear to have $S/N > 4$, even though they are simply random variation around the mean periodogram level. On the other hand, an excessively narrow interval may include part of the signal as noise, especially when red noise is present, decreasing the significance of the peak.
In this work, the noise level is computed using a moving median in log space with a filter width of equal to 0.2 in $\log 10$ space. 

\begin{table*}[th]
\caption{Main frequencies detected.}
\label{MainFreq}
        \centering
        \begin{tabular}{c|c|c|l}
 \hline \hline
                Object & Type & Sp.Typ. & Main frequencies \\
  \hline
                IGR J00370+6122 & SGXB & BN0.7Ib & none significant \\
                2S 0114+650 & SGXB & B1Iae & none significant \\
                Vela X-1 & SGXB & B0.5Iae & weak peak at 0.22 \cd\ (S/N: 5) \\
                Cen X-3 & SGXB & O6-7II-III & strong peak at 0.96 \cd\ (S/N: 26 - 38) \\
                1E 1145.1-6141 & SGXB & B2Iae & none significant \\
                4U 1538-522 & SGXB & B0.2Ia & peak at 0.535 \cd\ (S/N: 10) \\
                4U 1700-377 & SGXB & O6Ia & peak at 0.586 \cd\ (S/N: 6 - 10) \\
                Cyg X-1 & SGXB & O9.7Iab & weak peak at 0.36 \cd\ (S/N: 5 - 9) \\\hline
                IGR J08408-4503 & SFXT & O8.5Ib-II & weak peak at 0.27 \cd\ (S/N: 5 - 8) \\
                IGR J11215-5952 & SFXT & B1Ia & none significant \\
                IGR J16465-4507 & SFXT & B0.5-1Ib & very strong peak at 7.37 \cd\ + harmonics \\
                XTE J1739-302 & SFXT & O8Iab & none significant \\\hline
                RX J0146.9+6121 & BeXB & B1IIIe & frequency groups around 1.5 and 2.9 \cd\ \\
                X Per & BeXB & B1Ve & various groups around 1.2, 1.7, 2.88 and 3.6 \cd\ \\
                4U 1036-56 & BeXB & B0III-Ve & weak peak at 4.9 \cd\ (S/N: 5) \\
                4U 2206+543 & BeXB & O9.5Ve & weak peak at 6.15 \cd\ (S/N: 6) \\\hline
                XTE J0421+560 & sgB[e] & B0/2I[e] & strong peak at 2.46 \cd\ (S/N: 18-30) \\\hline
                4U 1954+319 & sg(M) & MI & main peak at 4.2 \cd\ with variable amplitude \\
        \hline \hline
 \end{tabular}
\end{table*}

\subsection{Light-curve gaps and interpolation}
\label{lcurves}

The obtained light curves are sampled at a constant frequency that depends on
the sector, but contain a variable amount of gaps. Some of the gaps are due to
planned outage in the measurements and others are due to invalid data in the
archive, such as those from excess scattered light. Because of the change in the TESS
operational strategy, light curves from sectors up to 55 have a large gap in the
middle of the sector of the order of 1 to 2 days. Subsequent sectors do not have
such large gaps, but have three gaps of around 5.25 hours every 6.85 days, which
is half the TESS orbital period. We performed a statistical study of the gaps
present in our light curves, finding that 70 light curves processed have 230
gaps of more than 1 hour, 70 gaps of more than 6 hours, and 50 gaps of more than
1 day.

The periodograms are computed using the Lomb-Scargle method, which in principle is designed to handle gaps and unevenly sampled data in time series.  However, we noticed that in four light curves  for sectors above 55, the computed periodogram showed strong artefacts linked to the periodic data gaps, which were not representative of the real periodogram of the source. This artificial effect was eliminated by performing a linear interpolation of the light curve over these short gaps, which introduces only a minor modification to the light curve. For consistency, the same process was followed for the light curves of all the sources.
We note that the filled gaps (less than 6 hours) are short compared to the characteristic timescales of the SLF variability being analyzed. In addition, only a few gaps per sector were filled, such that the total number of data points added by interpolation amounted to less than 4 percent of the overall dataset (i.e., less than 1 day over the 27 day sector).
Larger gaps were left unmodified as interpolating them would  have a greater  affect on the frequency content of the time series.


\subsection{Periodograms}

After detrending and pre-whitening, the final light curves were obtained and a
periodogram was computed for each sector.  The pre-whitened light curves consist
of stochastic signals and therefore their power spectrum is very noisy. A
standard method to reduce the noise in the periodograms is to perform averaging
in the frequency domain. In this  method, the data are divided up into segments
of equal duration, a power spectrum is obtained individually for each segment,
and then the resulting power spectra are  averaged. In this work, we average the
periodograms obtained in each sector to obtain a single mean periodogram per
source in addition to the individual periodograms per sector.

The time series are available at different sampling frequencies (see
Sect.~\ref{obs}), and therefore the maximum frequency sampled in the
periodograms (Nyquist frequency) is different for data obtained at different
epochs. However, as the averaging method requires all the periodograms to have
the same frequency range and resolution, we limited the frequency range of all
periodograms to the lowest Nyquist frequency, which is 24 d$^{-1}$. Before
computing the periodogram, and to account for the different white noise level
resulting from the different sampling frequency, the time series were re-sampled
at a  cadence of 30 minutes by averaging over three data points (sectors 27 to
55) or six data points (from sector 56).

While the 24 d$^{-1}$ frequency range is sufficient to analyze the SLF
variability, and even to extract higher-frequency peaks corresponding to
coherent pulsations, in particular in BeXBs, this limited frequency range made
the estimate of the white noise component less reliable, revealing significant
variability between sectors. In any case, as the parameters of interest are the
characteristics of the low-frequency noise component, this frequency range was
deemed acceptable within the scope of this work. This aspect is discussed in
more detail in Sect.~\ref{wn}.

\subsection{Noise-model fitting}
\label{model}

The present work focuses on the analysis of SLF variability in SGXB systems, and their comparison to BeXBs and isolated supergiants. The detection of low-frequency variability in a large sample of OB supergiant stars in \citet{bowman19} demonstrates the ubiquity of such phenomena in these stars. This type of low-frequency variability has been described in the literature as  ``red noise'', a type of correlated signal whose amplitude decreases as a function of frequency. This red noise is not due to measurement errors but is instead an intrinsic stochastic source of variability in the observed object. 

In the present work, we model the characterization of this variability in the frequency domain by fitting the amplitude spectrum $\alpha(f)$ of the normalized light curves with the function

\begin{equation}
 \alpha(f) = \frac{\alpha_0}{1 + \left(\frac{f}{f_c}\right)^{\gamma}} + \alpha_w
,\end{equation}

\noindent where $\alpha_0$ is the amplitude of the SLF variability at zero frequency, $f_c$ is its characteristic frequency, and $\gamma$ is the roll-off slope in logarithmic scale. The term $\alpha_w$ is used to model the white noise component of the spectrum, which is dominant at higher frequency. The characteristic frequency is the inverse of the characteristic timescale $f_c=(2\pi\tau)^{-1}$, where $\tau$ can be understood as the duration of the dominant structures in the light curve \citep{stanishev02,blomme11}.


This model has been used by many authors to characterize this type of
variability in various types of stars, for example for early-type stars in
\citet{bowman19, bowman19b, bowman20}, for yellow/red supergiants in
\citet{dorn-wallenstein20}, for Wolf-Rayet stars in \citet{lenoir22}, and for
luminous blue variables in \citet{dorn-wallenstein19} and \citet{naze21}

To fit this model to the periodogram, we employed the routine \texttt{curve\_fit} of  \textsc{SciPy}., which uses non-linear least squares to fit a given function to input data by minimizing either the sum of the square of the residuals or, if a standard deviation is provided for each data point, the $\chi^2$ value. Following \citet{bowman19b} and \citet{dorn-wallenstein19}, we performed the fit on the logarithm of the amplitude spectrum instead of performing it on the amplitude spectrum itself, as it is less sensitive to individual peaks in the periodogram.

To estimate the quality of the best-fit model, two indicators were computed: the
reduced chi squared $\chi^2_{\rm red}$ and the probability $Q$. A value of
$\chi^2_{\rm red}$ of close to unity indicates a good model fit, where the data points
are coherent with the model and the estimated error of the data. However, what
can be considered ``close'' depends on the number of degrees of freedom of the
fit. For this reason, the additional indicator $Q$, which gives the probability
that the chi-square will exceed a particular value $\chi^2$ by chance if the
model is correct, allows us to estimate the goodness of fit. If $Q$ is a very small probability for some particular data set, then the apparent discrepancies are unlikely to be chance fluctuations. In general, a value of $Q$ above 0.001 is acceptable \citep[chapter 15]{press93}.


\section{Results}

In this section, we present the results of our analysis. More details on
individual sources can be found in the Zenodo repository at
https://zenodo.org/uploads/14328326.

   \begin{figure*}
   \centering
     \includegraphics[width=16cm]{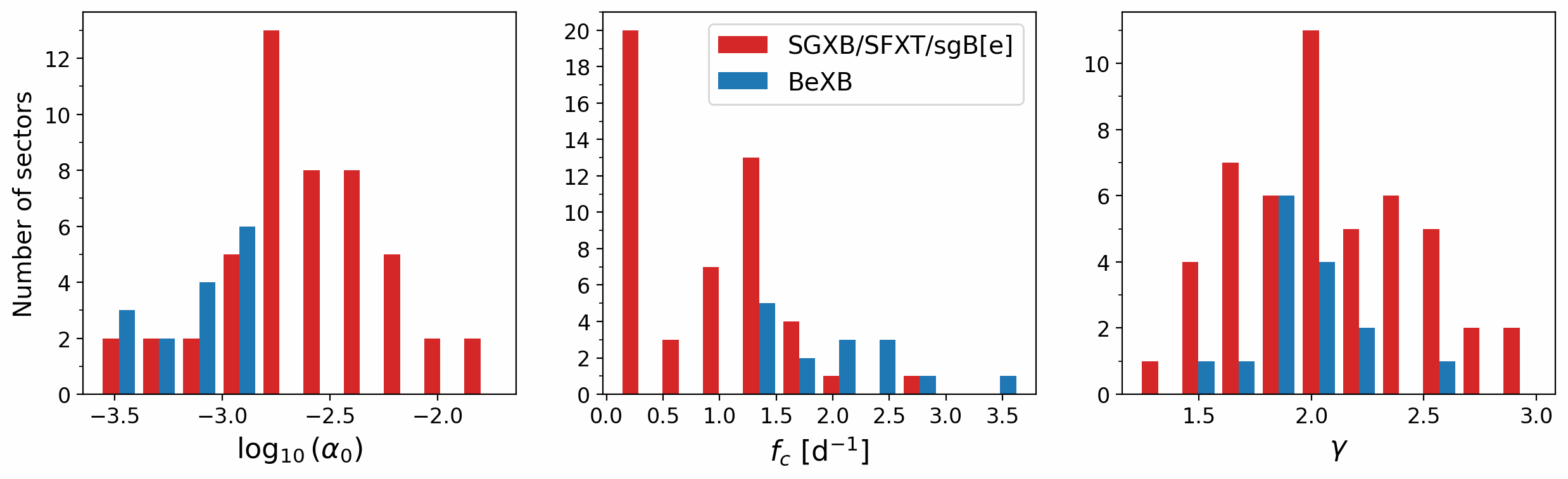} 
     \caption{SLF parameters: Histograms per sector and BeXB/SGXB.
      }
         \label{his:SGXB-BeXB}
   \end{figure*}

   \begin{figure}
   \centering
     \includegraphics[width=11cm]{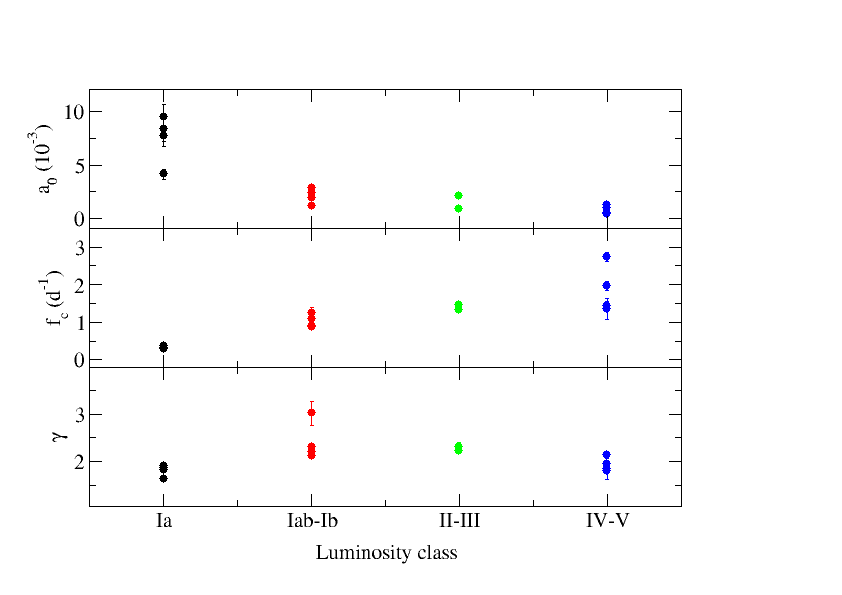}
     \caption{Best-fit parameters of the SLF variability as a function of luminosity class.}
         \label{par-lum}
   \end{figure}


\subsection{Significant frequencies}

As explained in Sect.~\ref{whitening}, prior to the computation of the periodogram, the light curves were searched for the presence of coherent signals. 
In general, the periodograms for all the analyzed SGXBs are featureless
above 2 \cd, with no significant peaks at high frequency, but instead a
progressive increase in the amplitude level towards low frequencies, in line with
 SLF variation in the form of a ``red noise'' on top of the
white noise floor level, which dominates at higher frequencies. 

Table~\ref{MainFreq} presents a summary of the presence or absence of
significant  frequencies for each object. The complete list of significant
frequencies detected, along with their amplitude and S/N, is provided in Appendix
\ref{sigfreq}.

\subsubsection{SGXBs}

Five of the SGXBs systems show a significant peak at low frequency,
corresponding to twice the orbital frequency of the system. The most obvious
example is Cen X-3, for which the light curves of the four sectors show a
clear pattern of double-wave ellipsoidal light variation, with two equal maxima
and two unequal minima per orbital period, which are produced by the revolution of a
tidally deformed star along its orbit \citep{vanparadijs83}. The peak in the periodogram is detected at 0.96 \cd, which is close to twice the orbital frequency
given the orbital period of 2.03 days. Slow periodic variation can be seen in
the light curves of 4U 1538-522 and 4U 1700-377 as well, but in this case the
exact pattern of the variation is less obvious as it is mixed with a higher level
of stochastic variability. In any case, the detected peaks in their periodograms
at 0.535 \cd and 0.586 \cd are equal to twice the orbital frequency, given
their respective orbital periods of 3.73 and 3.41 days, respectively. A similar
observation can be made in the periodograms of Cyg X-1 and Vela X-1, although in
these cases it is much harder to detect any clear periodic variation in the light
curves due to the longer periods involved. It should be highlighted that the
frequencies detected are identical in all the sectors of these sources.

The SFXT IGR J08408-4503 shows a low-amplitude peak at 0.27 \cd , which is also
close to twice the orbital frequency, given the orbital period of 9.5 days. We note that this peak is relatively small and is only detected in three out of five
sectors.

The only exception to this general behavior is seen in IGR J16465-4507, which is
classified as an SFXT, and shows $\beta$ Cephei-like pulsation with a
high-frequency peak at 7.37
\cd, with an amplitude of around 8 ppt, and harmonics of 14.75 \cd and 22.12
\cd. This source is discussed in more detail in Sect~\ref{disc:SGXB}.

Finally, the remaining SGXBs/SFXTs do not show a significant peak above the red
noise level. We point out that these systems all have longer orbital
periods, which would make detecting any existing peak more difficult, given the
27 day duration of the sectors.

\subsubsection{BeXBs}

X Persei has a large number of significant frequencies in various groups, which vary over time. The presence of such frequency groups is characteristic of Be stars, although not all pulsating Be stars display them \citep[e.g.,][]{baade18,reig22}. Sector 18 has mostly two groups, around 1.7 \cd\ and 3.6 \cd,
which also appear in sectors 43 and 44 in addition to an isolated peak at
2.88 \cd. On the other hand, sectors 70 and 71 have a group around 1.2 \cd\ and
another around 3.7 \cd. In addition, some sectors show peaks at higher
frequency, but with lower amplitude. 

RX J0146.9+6121 shows a fairly similar pattern to X Persei. In this case, the main
groups are around 1.5 \cd\ and 2.9 \cd\ and it also has some other peaks at higher
frequency with lower amplitude. This source was observed in only one sector, and so
it is not possible to assess any variability over time.

In the case of 4U 1036-56 and 4U 2206+543, the periodograms are quite different
as they do not contain a large number of significant frequencies, but instead
are mostly dominated by a red noise component at low frequency and a lower white noise component, similar to what is seen in SGXBs. However, contrary to
the SGXBs, there are some peaks at higher frequencies, which stand out above the
noise level. 

For 4U 2206+543, there is a peak at 6.15 \cd, which is detected in all sectors, with a
small amplitude of only 0.8 ppt. In the case of 4U 1036-56, a peak at 4.9 \cd\
is detected in sectors 10, 36, and 37 with an amplitude of around 1 ppt and a peak at
1.8 \cd\ is detected in sectors 10 and 37 with an amplitude of around 2 ppt. In
sectors 63 and 64, these peaks are below S/N = 4.

\subsubsection{Other sources}
The source XTE J0421+560 is the only one in the list classified as sgB[e]. This source shows some long-term variability, which is
more in line with BeXB sources than SGXBs. In terms of the periodograms, XTE J0421+560 shows
a strong peak at 2.46 \cd for all three sectors, and another peak at 1.09 \cd
in sectors 59 and 53, and in both cases the amplitude is around 6 ppt. 

The case of 4U 1954+319 is more peculiar, as it is a unique object. 4U 1954+31 is the only HMXB containing a late-type supergiant \citep{hinkle20}.
The periodograms show a peak on all sectors around 4.2 \cd; however, there is significant variation (by an order of magnitude) in the amplitude of the peak between sectors or even within a sector. 

   \begin{figure*}
   \centering
     \includegraphics[width=16cm]{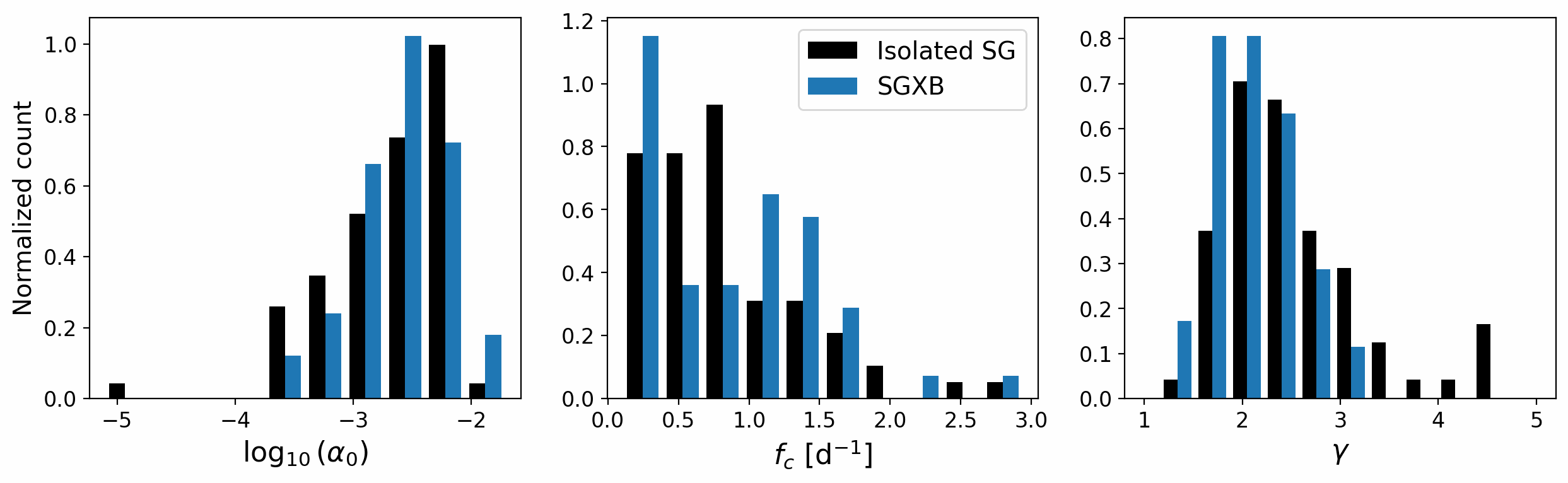} 
     \caption{Comparison of the best-fit SLF model parameters between SGXBs and single supergiant stars. Data for the isolated supergiants come from \citet{bowman19} and \citet{bowman20}.
      }
         \label{SGXB-SG}
   \end{figure*}


\subsection{Parameters of the noise model }
\label{noisepar}

After de-trending and pre-whitening, the final light curves were obtained and the
periodograms were  computed for each sector. The periodograms were fit with the model
described in Sect.~\ref{model}.  The noise-model parameters obtained for the
mean periodogram of each source are listed in Table \ref{tab3:modelParams}
along with indicators for the goodness of fit. The parameters obtained per
sector are provided in Appendix \ref{appendix:persector}. The normalized light
curves and the periodograms of the individual sources are available at
https://zenodo.org/uploads/14328326.

\subsubsection{Main results}

Figure~\ref{his:SGXB-BeXB} shows the histogram of the main model parameters for
SGXBs and BeXBs.  SGXBs display larger amplitude $a_0$ and smaller
characteristic frequencies $f_c$ than BeXBs. The roll-off slope, that is, how fast the
amplitude of the noise decreases as the frequency increases, follows a similar
distribution. In general, there is a correlation between the value of the
parameters and the overall luminosity, as can be seen in Fig.~\ref{par-lum}.

On the other hand, the range of values obtained for the red noise model
parameters are compatible with the ones obtained in the literature for the TESS sample of OB supergiants \citep{bowman19,bowman20}. Figure~\ref{SGXB-SG} compares the main model parameters obtained from our analysis with those of single supergiant stars.

We also investigated the stability of the parameters over time. This analysis was possible because most sources have been observed in more than one sector, with time intervals between observations of a few hundred days or even years. We find that, typically, all parameters except the white noise level, $\alpha_{w}$ , computed on each sector are within 3$\sigma$ of the parameter computed from the mean periodogram. In terms of the root mean square, which is computed as $rms=\sigma/av$, where  $av$ is the average value obtained for each parameter using each individual sector of a source and $\sigma$ is the standard deviation, the most stable parameter is $\gamma$ with $rms \simless 10$\%, followed by $f_c$ with $rms \simless 20$\%. For $a_0$, the rms varies in the range of 10--30\%. 



The white noise level is the parameter with the greatest dispersion. This parameter is discussed in the following section.

   \begin{figure*}
   \centering
   \begin{tabular}{cc}
     \includegraphics[width=9.5cm]{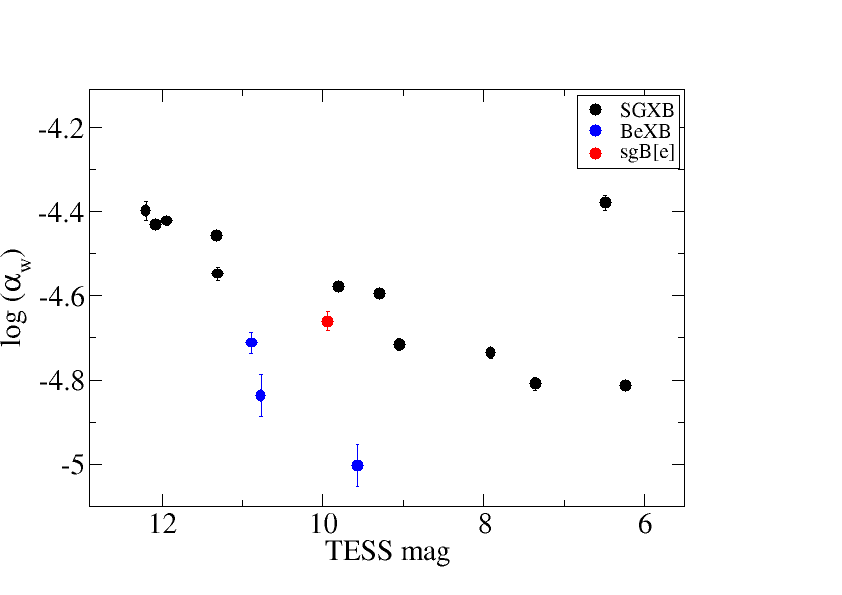}   &  
     \includegraphics[width=9.5cm]{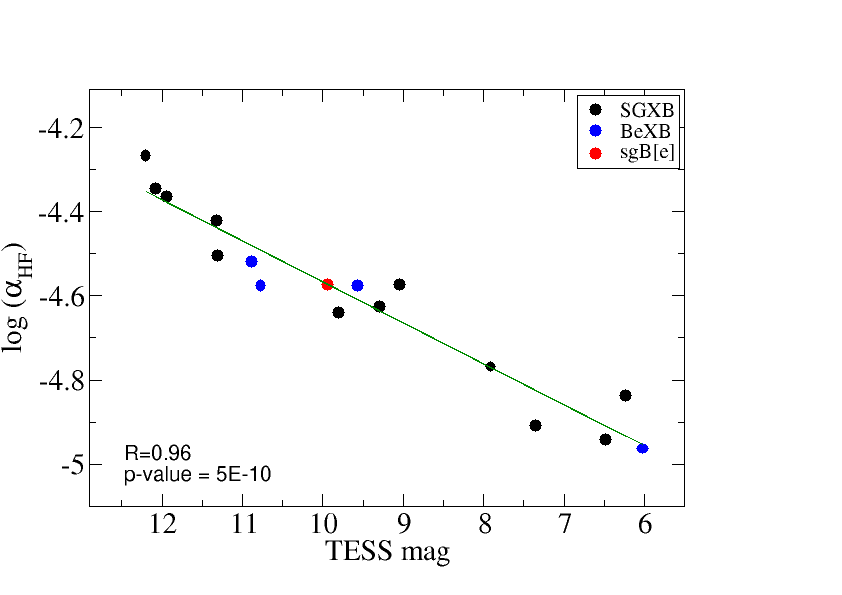}   \\
   \end{tabular}
     \caption{Correlation between the white noise computed from the periodogram (left) and that derived from the light curve (right) and the apparent magnitude. Blue and black points correspond to BeXBs and SGXBs, respectively. }
         \label{noise-mag}
   \end{figure*}


\subsubsection{White noise}
\label{wn}

There are two effects that influence the determination of the white noise level. First, high frequencies must be sampled to correctly define the noise level. Second, the intrinsic white noise is mixed with photon noise, which depends on the apparent magnitude and instrumental sensitivity \citep{bowman20,naze21}. 

As mentioned in Sect.~\ref{lcurves}, the time resolution of the TESS light curves changed over the course of the mission. The resolution determines the maximum frequency (Nyquist frequency) of the periodogram. To be able to use the maximum amount of data in a homogeneous way, we limited the maximum frequency to 24 d$^{-1}$ (time resolution of 1800 s). However, for sources with a
high value of $f_c$, as in the BeXBs, this maximum frequency proved to be insufficient, resulting in underestimated values of $\alpha_w$. This effect can be seen in the left panel of Fig.~\ref{noise-mag}, which shows the relationship between white noise and apparent magnitude. The left panel in this figure shows the white noise computed from the periodogram. The lower noise in BeXBs can be 
attributed to the limited range in frequencies that defines the noise level.
 To account for this effect, we computed the noise level directly from the light curves, as follows: the light curves were filtered by a high-pass filter with a cut-off frequency of 15 \cd to eliminate all the low-frequency components, and the variance of the residual signal was computed. This was then converted to an equivalent amplitude spectrum level $\alpha_{HF}$ , which is expected to be close to the value of $\alpha_w$ obtained through the periodogram fitting. Figure~\ref{noise-mag} illustrates the dependence of the white noise on the brightness of the source (i.e., photon noise). The straight line is simply the best log-linear fit to the data and shows that the correlation between white noise and brightness is statistically significant.


\begin{table*}
 \caption{Best-fit noise model parameters. }
        \label{tab3:modelParams}
        \centering
        \begin{tabular}{cccccccc}
 \hline\hline
                Source & Type & $\alpha_0 / 10^{-3}$ & $f_c$ [\cd] & $\gamma$ &  $\alpha_w / 10^{-5}$ & $\chi^2_{red}$ & $Q$ \\\hline         
        IGR J00370+6122 & SGXB & 2.90 $\pm$ 0.22 & 0.86 $\pm$ 0.06 & 2.12 $\pm$ 0.05 & 1.92 $\pm$ 0.06 & 1.20 & 3e-04 \\
        2S 0114+650 & SGXB & 9.48 $\pm$ 1.13 & 0.29 $\pm$ 0.03 & 1.81 $\pm$ 0.04 & 2.64 $\pm$ 0.07 & 0.93 & 0.910 \\
        Vela X-1 & SGXB & 7.68 $\pm$ 0.96 & 0.37 $\pm$ 0.05 & 1.63 $\pm$ 0.05 & 4.18 $\pm$ 0.17 & 1.02 & 0.340 \\
        Cen X-3 & SGXB & 0.90 $\pm$ 0.06 & 1.47 $\pm$ 0.09 & 2.31 $\pm$ 0.10 & 3.72 $\pm$ 0.08 & 1.26 & 8e-06 \\
        1E 1145.1-6141 & SGXB & 4.15 $\pm$ 0.48 & 0.33 $\pm$ 0.03 & 1.91 $\pm$ 0.05 & 3.50 $\pm$ 0.06 & 1.17 & 0.002 \\
        4U 1538-522 & SGXB & 1.92 $\pm$ 0.18 & 0.90 $\pm$ 0.07 & 2.30 $\pm$ 0.09 & 3.78 $\pm$ 0.09 & 1.03 & 0.310 \\
        4U 1700-377 & SGXB & 1.79 $\pm$ 0.12 & 1.58 $\pm$ 0.07 & 2.81 $\pm$ 0.08 & 1.53 $\pm$ 0.04 & 0.96 & 0.750 \\
        Cyg X-1 & SGXB & 2.37 $\pm$ 0.14 & 1.10 $\pm$ 0.06 & 2.21 $\pm$ 0.05 & 1.84 $\pm$ 0.05 & 1.09 & 0.056 \\
        \hline
        IGR J08408-4503 & SFXT & 2.12 $\pm$ 0.12 & 1.33 $\pm$ 0.07 & 2.21 $\pm$ 0.05 & 1.55 $\pm$ 0.06 & 1.08 & 0.085 \\
        IGR J11215-5952 & SFXT & 8.34 $\pm$ 1.12 & 0.29 $\pm$ 0.03 & 1.87 $\pm$ 0.04 & 2.54 $\pm$ 0.06 & 0.91 & 0.950 \\
        IGR J16465-4507 & SFXT & 0.39 $\pm$ 0.03 & 2.02 $\pm$ 0.22 & 1.81 $\pm$ 0.14 & 4.00 $\pm$ 0.20 & 1.05 & 0.200 \\
        XTE J1739-302 & SFXT & 1.24 $\pm$ 0.19 & 1.25 $\pm$ 0.14 & 3.02 $\pm$ 0.25 & 2.83 $\pm$ 0.10 & 1.11 & 0.030 \\
        \hline
        RX J0146.9+6121 & BeXB & 0.52 $\pm$ 0.09 & 1.35 $\pm$ 0.28 & 1.80 $\pm$ 0.18 & 1.46 $\pm$ 0.17 & 1.10 & 0.040 \\
        X Per & BeXB & 0.42 $\pm$ 0.02 & 2.73 $\pm$ 0.12 & 2.14 $\pm$ 0.07 & 0.72 $\pm$ 0.06 & 1.50 & 3e-15 \\
        4U 1036-56 & BeXB & 1.25 $\pm$ 0.08 & 1.44 $\pm$ 0.10 & 1.85 $\pm$ 0.06 & 1.94 $\pm$ 0.11 & 1.19 & 8e-04 \\
        4U 2206+543 & BeXB & 1.06 $\pm$ 0.06 & 1.97 $\pm$ 0.12 & 1.94 $\pm$ 0.06 & 0.99 $\pm$ 0.11 & 1.15 & 0.004 \\
        \hline
        XTE J0421+560 & sgB[e] & 3.43 $\pm$ 0.51 & 0.39 $\pm$ 0.06 & 1.57 $\pm$ 0.06 & 2.19 $\pm$ 0.11 & 1.09 & 0.051 \\
        \hline
        4U 1954+319 & sg(M) & 0.14 $\pm$ 0.01 & 1.66 $\pm$ 0.30 & 0.99 $\pm$ 0.09 & 0.43 $\pm$ 0.18 & 1.79 & 2e-31 \\
  \hline\hline
        \end{tabular}
    \tablefoot{Columns 3-6 give the amplitude of the SLF variability, its characteristic frequency, the roll-off slope in logarithmic scale, and the white noise level. Columns 7-8 provide the goodness of fit through the reduced $\chi^2$ and probability $Q$. Typically, a fit is considered to be good if $\chi^2_{\rm red} \approx 1$ or $Q>0.001$.}
\end{table*}

\section{Discussion}

The two main subcategories of HMXBs  are SGXBS and BeXBs, which can be
differentiated by the luminosity of the massive companion. This luminosity plays
a fundamental role in characterizing the fast optical variability in HMXBs.  The
aim of the present work is to investigate the optical variability  of SGXBs on
short timescales (hours to days) and to compare these systems with the related
BeXBs and with various types of single supergiant stars. We took advantage of
the rich data archive of the TESS mission and obtained light curves of a set of
SGXBs (see Table~\ref{list}). TESS light curves combine a fine time resolution
of a few tens of seconds and a duration of a few tens of days, with almost no
interruptions. Owing to the stochastic nature of the variability under
investigation, we performed our analysis in the frequency domain.

\subsection{Rapid optical variability of SGXBs: Main properties}
\label{sgxb:prop}

The rapid optical variability of SGXBs is characterized by stochastic noise and a lack of coherent pulsation modes. The amplitude spectrum is well represented by a phenomenological model that consists of a red noise component  that flattens at low frequencies. The red noise component extends up to the point where white noise dominates, typically up to $\sim 10$ \cd. However, the white noise level is dependent on the photon noise (i.e., brightness of the source) and the quality of the data (Fig.~\ref{noise-mag}).


A general trend of $a_0$ increasing and $f_c$ decreasing as the luminosity increases is observed (Fig.~\ref{par-lum}). In contrast, $\gamma$ does not exhibit any clear trend.

\subsection{SGXBs versus BeXBs}

In this work, we restricted the study of BeXBs to persistent (in X-rays) systems. A systematic study of BeXBs was presented in \citet{reig22}.

Although short-term variability is a common feature intrinsic to both SGXBs and
BeXBs, the morphology of the periodograms and the amplitude of the signal is
very different. In X-ray transient BeXBs, rapid variability  manifests as 
multi-periodic oscillations.  Many of the detected frequencies of the
modulations are higher than the maximum allowed for rotation, which favors the
interpretation that non-radial pulsations are the main driver of the fast
optical variability in BeXBs
\citep{reig22}. The periodograms of BeXBs are characterized by stochastic
variability on top of which coherent and quasi-periodic oscillations are
superimposed.  The most common pattern of variability in the periodograms of
BeXBs (as well as in classical Be stars) is the existence of frequency groups,
followed by isolated signals. Many transient BeXBs also display high-frequency
($>6$ \cd) signals, which can be interpreted as $\beta$ Cephei-like pulsations.
In contrast, the pulsational content in SGXBs is very scarce  and completely
absent above 2 \cd\ (Table~\ref{MainFreq}).

The stochastic variability between SGXBs and BeXBs is also different. 
While in SGXBs, SLF variability is characterized by red noise and flattens toward lower frequencies, in BeXBs the low-frequency part of the periodogram also exhibits a red-noise-like component, but in this case, it is made up of a forest of signals whose origin is believed to be inhomogeneities in the inner circumstellar disk and/or light outbursts \citep{reig22}. The differences between SGXBs and BeXBs are also notorious even when considering persistent BeXBs, as can be seen in Fig.~\ref{par-lum}.

\subsection{SGXBs versus single supergiant stars}
\label{disc:SGXB}

Similarly to the relationship between BeXBs and classical Be stars, the rapid
optical variability of  SGXBs  is very similar to that of isolated supergiant
stars. The range of values obtained for the red noise model parameters is
compatible with that found in the literature for isolated supergiants
\citep{bowman19,bowman20,dorn-wallenstein20,naze21}. In particular,  the
logarithmic amplitude gradient $\gamma$ varies typically in a range between 1.5
and 3.5, with a higher concentration around $\gamma=2$, while the characteristic
frequency  is distributed mainly below 1.5 \cd\ (see Fig.~\ref{SGXB-SG}).
Likewise, the dependence between the noise parameters and the luminosity
described in Sect.~\ref{sgxb:prop} also agrees with the results of
\citet{bowman19}, who found a similar correlation in their TESS sample of
supergiant stars in the Large Magellanic Cloud. 

The lack of significant frequencies among SGXBs is in line with the results presented in \citet{bowman19}, where among the type Ia  supergiants all but one show a frequency spectrum containing
mainly red noise without any other peaks. Among the sample of 91 supergiants, only two show $\beta$ Cephei-like pulsations, and both  are type Ib. 

In our sample of SGXBs, only IGR J16465-4507 shows a $\beta$ Cephei-like pulsation.  IGR J16465-4507 was classified as type B0.5-1Ib
by \citet{chaty16}, who also obtained the values of \top = 26\,000 K and \gop =
3.1 from spectral modeling. These parameters place the star within the
boundaries of the theoretical $\beta$ Cephei instability strip
\citep{pamyatnykh99}. Another peculiarity of this source is its high rotational
speed, estimated at 320 km s$^{-1}$ \citep{chaty16}. This value is significantly
higher than typical values in SGXBs \citep{liu06}. Moreover, H$\alpha$ emission
is also observed, which led \citet{chaty16} to speculate that it may come from
circumstellar material, possibly with a disk-like geometry.  Finally, its
position on the Corbet diagram places it between Be and supergiant systems, and
therefore it is possibly a descendant of BeXBs. In view of these facts, it is
not surprising that our analysis reveals that the best-fit noise model
parameters of IGR J16465-4507 are more aligned with those of BeXBs than those of
SGXBs.  

\subsection{Origin of the variability}

The physical cause of the SLF variability in early-type massive stars is 
unclear. Internal gravity waves (IGWs) excited by core convection have been
proposed to explain this low-frequency variability
\citep{rogers13,aerts15,edelmann19,bowman19,ratnasingam20,ratnasingam23,vanon23,herwig23,thompson24}.
However, other authors, such as \citet{lecoanet19} and \citet{cantiello21},
argue that the observed low-frequency variability in massive stars is unlikely
to be due to internal gravity waves generated by core convection, because they
would suffer strong radiative damping toward the surface, resulting in very
little surface luminosity variation \citep[see also][]{anders23}. Instead, 
these authors attribute
the presence of red noise to  subsurface convection zones driven by the iron
opacity peak.

The models disagree even in the interpretation of the observations.
\citet{bowman19} claimed that the correlation between $\alpha_0$ and brightness
supports the interpretation that the SLF variability observed in OB stars is
evidence of IGWs, arguing that the amplitudes of individual waves caused by core
convection are predicted to scale with the local core luminosity, and therefore
the mass of the convective core. In contrast, \citet{cantiello21} argued that
this correlation should go in the opposite direction, because IGWs are expected
to have a weaker influence toward higher luminosities. Also, the predicted
characteristic frequencies expected to result from IGWs  are significantly
smaller than those observed \citep{shiode13,cantiello21}. Finally, the
correlation between the characteristic frequency and luminosity
(Fig.~\ref{par-lum}) would also be at odds with the SLF variability being due to
IGWs  because the characteristic frequencies of peak IGWs should increase as
stars evolve on the main sequence \citet{cantiello21}, which is the opposite of
what is observed.

Owing to the complexity of the problem, analytical models make some
simplifying assumptions (e.g., mixing-length theory, non-rotating models, linear
wave propagation, and no wave--wave interaction), which limit our ability to make accurate comparisons to observations, especially for massive evolved stars. Studies of the excitation and propagation of gravity waves have greatly benefited
from 3D hydrodynamic simulations in recent years, which allow the inclusion of those
simplifying assumptions, and coverage of a significant fraction of the star's
radius, but these simulations have not solved the controversy.  Many
hydrodynamic simulations appear to validate the effectiveness of IGWs in
explaining the SLF variability and are qualitatively consistent with
observations \citep{edelmann19,ratnasingam23,vanon23,herwig23,thompson24}. In contrast, other models that simulate  gravity
waves driven by core convection predict that the photometric variability due to those waves is orders of magnitude lower than the observed variability \citep{anders23}.

In either case, these models remain to be applied to more evolved stars
(i.e., supergiants). From stellar evolution theory, it is expected that the size
of the subsurface convection zone grows  as stars evolve. Hence, the contribution to
SLF variability from subsurface convection zones may be the dominant process in
supergiant stars \citep{bowman22}. On the other hand, the role of
inhomogeneities in the stellar surface or in the wind, in combination with
rotation \citep{aerts17,simon18,krticka21}, must also be taken into account.

\section{Conclusion}

We studied the rapid optical variability of the massive counterparts in SGXBs.
We show that most of the power in the periodograms of SGXBs is not caused by
pulsation but by SLF variability.  When this variability is described by a
phenomenological model, the parameters of the model that describe the amplitude
at zero frequency $a_0$ and the logarithmic amplitude gradient (i.e., the slope
of the linear decrease) $\gamma$ appear to depend on luminosity
(Fig.~\ref{par-lum}). More luminous sources tend to show variability of greater
amplitude and lower logarithmic amplitude  gradients than less luminous sources. The
statistical significance of this trend should be assessed when reliable values
of the luminosity become available.  The specific physical cause of this noise
remains unclear. Such low-frequency variability could be produced by a
combination of internal gravity waves excited at the interface between the
convective core and the radiative envelope or in a subsurface convection zone.

The amplitude spectra of SGXBs and BeXBs display very different components. In terms of the pulsational content, the main difference is the disappearance of coherent signals as the luminosity increases. In terms of the SLF variability, in SGXBs, the amplitude $a_0$ and the characteristic frequency $f_c$ (i.e., the characteristic variability timescale) decrease as the luminosity increases.

\section{Data availability}

The light curves and periodograms of individual sources are available on the Zenodo
repository: https://zenodo.org/uploads/14328326

\begin{acknowledgements} 

We thank the anonymous referee for his/her useful comments. This paper includes data collected by the TESS mission. Funding for the TESS mission is provided by the NASA's Science Mission Directorate. This research has made use of the SIMBAD database, operated at CDS, Strasbourg, France and of NASA’s Astrophysics Data System operated by the Smithsonian Astrophysical Observatory. 

\end{acknowledgements}

\bibliographystyle{aa}
\bibliography{./artBex_bib}



\begin{appendix}

\section{Significant independent frequencies}
\label{sigfreq}

Independent frequencies of the different types of sources. 

\begin{table}[ht]
\caption{Significant frequencies - SGXBs.}
\label{freq-sgxb}
\centering
\begin{tabular}{lccc}
   \hline       
        & Frequency & Amplitude & SNR \\
   & [\cd] & [ppt] & \\
\hline
\multicolumn{4}{l}{ Vela X-1  } \\
Sector 8 & 0.220 & 32.48 & 5.1 \\
 & 0.122 & 16.59 & 5.2 \\
Sector 9 & 0.217 & 26.63 & 5.9 \\
 & 0.174 & 12.12 & 4.6 \\
Sector 35 & None & & \\
Sector 62 & 0.225 & 31.37 & 5.2 \\
 & 0.133 & 15.90 & 4.0 \\
\hline
\multicolumn{4}{l}{ Cen X-3  } \\
Sector 10 & 0.959 & 21.98 & 26.3 \\
 & 1.435 & 2.95 & 6.0 \\
Sector 11 & 0.957 & 20.85 & 38.8 \\
 & 1.440 & 2.80 & 6.6 \\
 & 0.484 & 3.77 & 4.8 \\
Sector 37 & 0.958 & 17.69 & 28.3 \\
 & 1.437 & 2.67 & 7.2 \\
Sector 64 & 0.958 & 18.25 & 26.1 \\
 & 1.428 & 1.97 & 4.5 \\
 & 3.834 & 0.62 & 4.3 \\
\hline
\multicolumn{4}{l}{ 4U 1538-522  } \\
Sector 12 & 0.536 & 12.16 & 10.0 \\
Sector 39 & 0.533 & 9.99 & 11.4 \\
Sector 65 & 0.540 & 11.83 & 10.9 \\
\hline
\multicolumn{4}{l}{ 4U 1700-377  } \\
Sector 12 & 0.589 & 13.78 & 10.0 \\
Sector 39 & 0.587 & 15.44 & 10.3 \\
Sector 66 & 0.584 & 13.32 & 6.2 \\
\hline
\multicolumn{4}{l}{ Cyg X-1  } \\
Sector 14 & 0.371 & 14.46 & 5.4 \\
Sector 54 & 0.356 & 17.27 & 9.4 \\
 & 0.188 & 8.23 & 5.0 \\
Sector 55 & 0.359 & 17.08 & 9.5 \\
Sector 74 & 0.355 & 20.98 & 12.6 \\
Sector 75 & 0.371 & 13.09 & 5.2 \\
 & 2.337 & 1.46 & 4.8 \\
\hline
\multicolumn{4}{l}{ IGR J08408-4503  } \\
Sector 8 & None & & \\
Sector 9 & None & & \\
Sector 35 & 0.271 & 13.49 & 5.0 \\
Sector 61 & 0.275 & 11.90 & 5.0 \\
Sector 62 & 0.275 & 9.87 & 7.8 \\
\hline
\multicolumn{4}{l}{ IGR J16465-4507  } \\
Sector 12 & 7.374 & 7.22 & 74.4 \\
 & 14.750 & 0.78 & 10.4 \\
Sector 39 & 7.374 & 9.35 & 114.1 \\
 & 14.749 & 1.08 & 19.9 \\
 & 22.131 & 0.27 & 5.4 \\
Sector 66 & 7.374 & 7.16 & 88.2 \\
 & 14.748 & 0.83 & 16.2 \\
 & 22.123 & 0.25 & 5.7 \\
 \hline 
\end{tabular}
\end{table}

\begin{table}
\caption{Significant frequencies - BeXBs. }
\label{freq-bexb}
\centering
\begin{tabular}{lccc}
\hline  
   & Frequency & Amplitude & SNR \\
   & [\cd] & [ppt] & \\
\hline
\multicolumn{4}{l}{ RX J0146.9+6121  } \\
Sector 58 & 9.680 & 1.53 & 35.6 \\
 & 13.293 & 0.64 & 22.1 \\
 & 14.282 & 0.35 & 12.5 \\
 & 2.923 & 1.92 & 12.8 \\
 & 8.417 & 0.32 & 8.2 \\
 & 22.966 & 0.14 & 6.3 \\
 & 15.265 & 0.16 & 5.8 \\
 & 1.502 & 1.97 & 5.8 \\
 & 16.719 & 0.12 & 5.1 \\
 & 2.885 & 0.67 & 5.0 \\
 & 17.884 & 0.11 & 4.5 \\
\hline
\multicolumn{4}{l}{ 4U 1036-56  } \\
Sector 10 & 4.912 & 0.85 & 5.3 \\
 & 5.647 & 0.50 & 4.3 \\
 & 1.851 & 1.85 & 4.1 \\
Sector 36 & 4.910 & 0.77 & 4.3 \\
Sector 37 & 4.917 & 1.07 & 5.7 \\
 & 1.874 & 2.70 & 5.6 \\
\hline
\multicolumn{4}{l}{ 4U 2206+543  } \\
Sector 16 & 6.143 & 0.78 & 5.8 \\
Sector 17 & 6.136 & 0.86 & 6.1 \\
Sector 56 & 6.138 & 0.72 & 5.2 \\
Sector 57 & 6.135 & 0.88 & 6.5 \\
\hline
\multicolumn{4}{l}{ X Per  } \\
Sector 18 & 3.607 & 1.90 & 9.0 \\
 & 1.742 & 2.74 & 10.0 \\
 & 8.626 & 0.38 & 8.2 \\
 & 3.686 & 1.27 & 6.8 \\
 & 10.410 & 0.20 & 5.8 \\
 & 1.923 & 1.12 & 4.4 \\
 & 3.841 & 0.79 & 4.5 \\
Sector 43 & 1.738 & 3.42 & 9.0 \\
 & 3.614 & 2.14 & 8.9 \\
 & 2.885 & 1.88 & 7.0 \\
 & 3.679 & 1.50 & 7.0 \\
 & 3.836 & 1.12 & 5.4 \\
 & 10.368 & 0.15 & 4.3 \\
 & 8.635 & 0.21 & 4.3 \\
Sector 44 & 3.612 & 3.72 & 15.1 \\
 & 2.886 & 2.00 & 6.8 \\
 & 1.832 & 2.49 & 5.7 \\
 & 1.723 & 2.43 & 5.7 \\
 & 3.837 & 0.95 & 4.2 \\
 & 3.664 & 0.90 & 4.2 \\
Sector 70 & 6.793 & 0.70 & 15.5 \\
 & 13.588 & 0.21 & 12.2 \\
 & 1.216 & 2.26 & 12.8 \\
 & 3.526 & 1.02 & 9.2 \\
 & 3.634 & 0.93 & 8.5 \\
 & 20.383 & 0.08 & 7.8 \\
 & 3.679 & 0.78 & 7.0 \\
 & 3.741 & 0.65 & 5.9 \\
 & 6.497 & 0.25 & 5.6 \\
 & 7.092 & 0.24 & 5.6 \\
 & 3.815 & 0.57 & 5.4 \\
 & 11.592 & 0.10 & 4.5 \\
Sector 71 & 3.623 & 1.71 & 12.2 \\
 & 3.665 & 1.11 & 8.1 \\
 & 1.218 & 1.98 & 10.5 \\
 & 3.841 & 0.89 & 6.9 \\
 & 3.503 & 0.76 & 5.9 \\
 & 3.731 & 0.64 & 5.2 \\
 & 7.086 & 0.19 & 4.3 \\
 \hline 
\end{tabular}
\end{table}

\begin{table}
\caption{Significant frequencies - Others.}
        \centering
        \begin{tabular}{lccc}
   & Frequency & Amplitude & SNR \\
   & [\cd] & [ppt] & \\
\hline
\multicolumn{4}{l}{ XTE J0421+560  } \\
Sector 19 & 2.466 & 7.37 & 28.8 \\
Sector 59 & 2.466 & 6.64 & 29.6 \\
 & 1.097 & 5.34 & 10.9 \\
Sector 73 & 2.467 & 5.41 & 17.7 \\
 & 1.096 & 5.88 & 10.7 \\
\hline
\multicolumn{4}{l}{ 4U 1954+319  } \\
Sector 14 & 4.354 & 0.30 & 10.0 \\
 & 12.104 & 0.10 & 4.4 \\
Sector 41 & 4.258 & 3.04 & 26.4 \\
 & 8.515 & 1.14 & 15.9 \\
 & 4.205 & 0.82 & 7.2 \\
 & 12.752 & 0.24 & 5.3 \\
 & 4.324 & 0.55 & 5.0 \\
 & 8.578 & 0.33 & 4.7 \\
 & 4.405 & 0.49 & 4.5 \\
 & 17.495 & 0.15 & 4.1 \\
Sector 54 & 4.201 & 0.32 & 5.8 \\
Sector 55 & 4.195 & 0.37 & 18.1 \\
 & 8.381 & 0.09 & 6.9 \\
 & 4.238 & 0.13 & 6.3 \\
 & 4.284 & 0.10 & 4.8 \\
 & 8.440 & 0.05 & 4.1 \\
Sector 74 & 4.202 & 0.70 & 22.2 \\
 & 8.407 & 0.33 & 19.2 \\
 & 8.441 & 0.20 & 11.8 \\
 & 4.236 & 0.31 & 10.7 \\
 & 8.364 & 0.13 & 8.0 \\
 & 12.625 & 0.06 & 5.6 \\
Sector 75 & 4.247 & 0.96 & 33.6 \\
 & 8.492 & 0.54 & 32.1 \\
 & 21.250 & 0.09 & 9.9 \\
 & 12.743 & 0.09 & 7.7 \\
 & 8.524 & 0.12 & 7.5 \\
 & 16.979 & 0.06 & 6.2 \\
 \hline 
        \end{tabular}
\end{table}

\newpage

\section{Results per sector}
\label{appendix:persector}

The following table presents the best-fit noise model obtained for each sector
of each source as well as for its mean periodogram, grouped by source type.

\begin{table*}[ht]
\caption{Noise model parameters - SGXBs.}
\label{tab3:sectorModelParamsSGXB}
\centering
\begin{tabular}{c|c|c|c|c|c|c}
Source & Sector & $\alpha_0 / 10^{-3}$ & $f_c$ [\cd] & $\gamma$ &  $\alpha_w / 10^{-5}$ & $\chi^2_{red}$ \\
\hline  \hline
IGR J00370+6122 & 17 & 3.75 $\pm$ 0.79 & 0.68 $\pm$ 0.12 & 2.06 $\pm$ 0.12 & 2.19 $\pm$ 0.15 & 1.08 \\
IGR J00370+6122 & 18 & 2.33 $\pm$ 0.43 & 0.92 $\pm$ 0.15 & 2.10 $\pm$ 0.14 & 2.03 $\pm$ 0.15 & 1.04 \\
IGR J00370+6122 & 24 & 1.92 $\pm$ 0.34 & 1.00 $\pm$ 0.16 & 2.14 $\pm$ 0.14 & 1.75 $\pm$ 0.13 & 0.93 \\
IGR J00370+6122 & 58 & 2.12 $\pm$ 0.36 & 1.00 $\pm$ 0.13 & 2.31 $\pm$ 0.12 & 0.96 $\pm$ 0.07 & 0.95 \\
IGR J00370+6122 & Mean & 2.90 $\pm$ 0.22 & 0.86 $\pm$ 0.06 & 2.12 $\pm$ 0.05 & 1.92 $\pm$ 0.06 & 1.20 \\ 
\hline
2S 0114+650 & 18 & 13.24 $\pm$ 5.74 & 0.17 $\pm$ 0.07 & 1.57 $\pm$ 0.09 & 2.49 $\pm$ 0.24 & 0.93 \\
2S 0114+650 & 24 & 8.99 $\pm$ 2.67 & 0.33 $\pm$ 0.08 & 1.94 $\pm$ 0.11 & 2.74 $\pm$ 0.16 & 1.10 \\
2S 0114+650 & 25 & 5.24 $\pm$ 1.50 & 0.37 $\pm$ 0.09 & 1.86 $\pm$ 0.11 & 1.85 $\pm$ 0.13 & 0.90 \\
2S 0114+650 & 52 & 6.44 $\pm$ 2.08 & 0.30 $\pm$ 0.08 & 1.76 $\pm$ 0.10 & 1.87 $\pm$ 0.14 & 0.87 \\
2S 0114+650 & 58 & 8.53 $\pm$ 2.45 & 0.35 $\pm$ 0.08 & 2.06 $\pm$ 0.12 & 2.96 $\pm$ 0.15 & 0.32 \\
2S 0114+650 & Mean & 9.48 $\pm$ 1.13 & 0.29 $\pm$ 0.03 & 1.81 $\pm$ 0.04 & 2.64 $\pm$ 0.07 & 0.93 \\ 
\hline
Vela X-1 & 8 & 6.84 $\pm$ 2.18 & 0.38 $\pm$ 0.14 & 1.46 $\pm$ 0.12 & 5.45 $\pm$ 0.70 & 0.92 \\
Vela X-1 & 9 & 4.66 $\pm$ 1.36 & 0.43 $\pm$ 0.13 & 1.54 $\pm$ 0.12 & 3.06 $\pm$ 0.38 & 0.80 \\
Vela X-1 & 35 & 7.81 $\pm$ 2.25 & 0.38 $\pm$ 0.10 & 1.76 $\pm$ 0.11 & 3.53 $\pm$ 0.28 & 1.01 \\
Vela X-1 & 62 & 7.54 $\pm$ 2.21 & 0.36 $\pm$ 0.09 & 1.81 $\pm$ 0.10 & 2.46 $\pm$ 0.19 & 0.46 \\
Vela X-1 & Mean & 7.68 $\pm$ 0.96 & 0.37 $\pm$ 0.05 & 1.63 $\pm$ 0.05 & 4.18 $\pm$ 0.17 & 1.02 \\ 
\hline
Cen X-3 & 10 & 0.98 $\pm$ 0.18 & 1.18 $\pm$ 0.22 & 2.09 $\pm$ 0.21 & 3.57 $\pm$ 0.21 & 0.91 \\
Cen X-3 & 11 & 0.60 $\pm$ 0.08 & 2.12 $\pm$ 0.25 & 2.72 $\pm$ 0.29 & 3.53 $\pm$ 0.17 & 1.08 \\
Cen X-3 & 37 & 0.75 $\pm$ 0.12 & 1.36 $\pm$ 0.22 & 2.23 $\pm$ 0.21 & 2.65 $\pm$ 0.15 & 0.98 \\
Cen X-3 & 64 & 0.72 $\pm$ 0.10 & 1.68 $\pm$ 0.22 & 2.64 $\pm$ 0.27 & 3.47 $\pm$ 0.16 & 1.03 \\
Cen X-3 & Mean & 0.90 $\pm$ 0.06 & 1.47 $\pm$ 0.09 & 2.31 $\pm$ 0.10 & 3.72 $\pm$ 0.08 & 1.26 \\ 
\hline
1E 1145.1-6141 & 10 & 3.08 $\pm$ 0.87 & 0.39 $\pm$ 0.10 & 2.10 $\pm$ 0.18 & 3.69 $\pm$ 0.15 & 1.01 \\
1E 1145.1-6141 & 11 & 3.62 $\pm$ 1.05 & 0.35 $\pm$ 0.08 & 2.11 $\pm$ 0.15 & 2.48 $\pm$ 0.10 & 1.14 \\
1E 1145.1-6141 & 37 & 3.28 $\pm$ 1.07 & 0.31 $\pm$ 0.10 & 1.75 $\pm$ 0.13 & 2.44 $\pm$ 0.14 & 0.86 \\
1E 1145.1-6141 & 38 & 4.33 $\pm$ 1.44 & 0.29 $\pm$ 0.09 & 1.87 $\pm$ 0.14 & 3.30 $\pm$ 0.15 & 1.02 \\
1E 1145.1-6141 & 64 & 4.56 $\pm$ 1.55 & 0.28 $\pm$ 0.09 & 1.80 $\pm$ 0.14 & 3.36 $\pm$ 0.17 & 0.96 \\
1E 1145.1-6141 & Mean & 4.15 $\pm$ 0.48 & 0.33 $\pm$ 0.03 & 1.91 $\pm$ 0.05 & 3.50 $\pm$ 0.06 & 1.17 \\ 
\hline
4U 1538-522 & 12 & 1.31 $\pm$ 0.21 & 1.18 $\pm$ 0.16 & 2.62 $\pm$ 0.23 & 3.60 $\pm$ 0.15 & 1.16 \\
4U 1538-522 & 39 & 1.64 $\pm$ 0.30 & 0.93 $\pm$ 0.15 & 2.32 $\pm$ 0.19 & 3.33 $\pm$ 0.16 & 1.06 \\
4U 1538-522 & 65 & 1.97 $\pm$ 0.40 & 0.79 $\pm$ 0.14 & 2.16 $\pm$ 0.17 & 3.05 $\pm$ 0.16 & 0.98 \\
4U 1538-522 & Mean & 1.92 $\pm$ 0.18 & 0.90 $\pm$ 0.07 & 2.30 $\pm$ 0.09 & 3.78 $\pm$ 0.09 & 1.03 \\ 
\hline
4U 1700-377 & 12 & 1.41 $\pm$ 0.19 & 1.62 $\pm$ 0.17 & 2.75 $\pm$ 0.17 & 1.75 $\pm$ 0.10 & 0.90 \\
4U 1700-377 & 39 & 1.49 $\pm$ 0.19 & 1.72 $\pm$ 0.15 & 2.99 $\pm$ 0.17 & 1.43 $\pm$ 0.07 & 0.92 \\
4U 1700-377 & 66 & 2.04 $\pm$ 0.29 & 1.31 $\pm$ 0.14 & 2.64 $\pm$ 0.13 & 1.07 $\pm$ 0.07 & 0.79 \\
4U 1700-377 & Mean & 1.79 $\pm$ 0.12 & 1.58 $\pm$ 0.07 & 2.81 $\pm$ 0.08 & 1.53 $\pm$ 0.04 & 0.96 \\ 
\hline
Cyg X-1 & 14 & 2.32 $\pm$ 0.44 & 0.92 $\pm$ 0.16 & 1.96 $\pm$ 0.13 & 1.91 $\pm$ 0.19 & 1.08 \\
Cyg X-1 & 54 & 1.87 $\pm$ 0.27 & 1.33 $\pm$ 0.15 & 2.50 $\pm$ 0.15 & 1.58 $\pm$ 0.10 & 1.03 \\
Cyg X-1 & 55 & 2.06 $\pm$ 0.32 & 1.17 $\pm$ 0.14 & 2.42 $\pm$ 0.13 & 1.07 $\pm$ 0.08 & 1.01 \\
Cyg X-1 & 74 & 1.78 $\pm$ 0.28 & 1.23 $\pm$ 0.17 & 2.23 $\pm$ 0.14 & 1.62 $\pm$ 0.13 & 0.88 \\
Cyg X-1 & 75 & 2.13 $\pm$ 0.38 & 1.01 $\pm$ 0.17 & 2.03 $\pm$ 0.13 & 1.81 $\pm$ 0.17 & 0.64 \\
Cyg X-1 & Mean & 2.37 $\pm$ 0.14 & 1.10 $\pm$ 0.06 & 2.21 $\pm$ 0.05 & 1.84 $\pm$ 0.05 & 1.09 \\ 
\hline \hline
\end{tabular}
\end{table*}

\begin{table*}[h]
\caption{Noise model parameters - SFXTs.}
\label{tab3:sectorModelParamsSFXT}
\centering
\begin{tabular}{c|c|c|c|c|c|c}
Source & Sector & $\alpha_0 / 10^{-3}$ & $f_c$ [\cd] & $\gamma$ &  $\alpha_w / 10^{-5}$ & $\chi^2_{red}$ \\
\hline  \hline
IGR J08408-4503 & 8 & 1.98 $\pm$ 0.33 & 1.20 $\pm$ 0.19 & 2.03 $\pm$ 0.14 & 1.95 $\pm$ 0.21 & 0.91 \\
IGR J08408-4503 & 9 & 1.36 $\pm$ 0.18 & 1.63 $\pm$ 0.18 & 2.54 $\pm$ 0.16 & 1.48 $\pm$ 0.10 & 1.07 \\
IGR J08408-4503 & 35 & 2.80 $\pm$ 0.49 & 1.05 $\pm$ 0.17 & 2.05 $\pm$ 0.13 & 2.03 $\pm$ 0.21 & 0.87 \\
IGR J08408-4503 & 61 & 1.70 $\pm$ 0.24 & 1.45 $\pm$ 0.16 & 2.33 $\pm$ 0.11 & 0.41 $\pm$ 0.07 & 0.88 \\
IGR J08408-4503 & 62 & 1.53 $\pm$ 0.22 & 1.47 $\pm$ 0.18 & 2.29 $\pm$ 0.14 & 1.18 $\pm$ 0.12 & 0.75 \\
IGR J08408-4503 & Mean & 2.12 $\pm$ 0.12 & 1.33 $\pm$ 0.07 & 2.21 $\pm$ 0.05 & 1.55 $\pm$ 0.06 & 1.08 \\ 
\hline
IGR J11215-5952 & 10 & 4.96 $\pm$ 1.46 & 0.36 $\pm$ 0.09 & 1.81 $\pm$ 0.11 & 2.33 $\pm$ 0.16 & 0.94 \\
IGR J11215-5952 & 11 & 19.52 $\pm$ 11.17 & 0.10 $\pm$ 0.05 & 1.60 $\pm$ 0.09 & 1.91 $\pm$ 0.15 & 0.65 \\
IGR J11215-5952 & 37 & 7.63 $\pm$ 2.35 & 0.30 $\pm$ 0.07 & 2.01 $\pm$ 0.11 & 1.75 $\pm$ 0.09 & 0.89 \\
IGR J11215-5952 & 64 & 4.17 $\pm$ 0.89 & 0.63 $\pm$ 0.10 & 2.38 $\pm$ 0.16 & 3.22 $\pm$ 0.14 & 0.98 \\
IGR J11215-5952 & Mean & 8.34 $\pm$ 1.12 & 0.29 $\pm$ 0.03 & 1.87 $\pm$ 0.04 & 2.54 $\pm$ 0.06 & 0.91 \\ 
\hline
IGR J16465-4507 & 12 & 0.53 $\pm$ 0.11 & 1.40 $\pm$ 0.36 & 1.74 $\pm$ 0.26 & 4.41 $\pm$ 0.37 & 1.05 \\
IGR J16465-4507 & 39 & 0.27 $\pm$ 0.04 & 2.94 $\pm$ 0.43 & 2.39 $\pm$ 0.38 & 3.88 $\pm$ 0.27 & 1.03 \\
IGR J16465-4507 & 66 & 0.36 $\pm$ 0.09 & 1.45 $\pm$ 0.58 & 1.23 $\pm$ 0.24 & 1.81 $\pm$ 0.61 & 0.94 \\
IGR J16465-4507 & Mean & 0.39 $\pm$ 0.03 & 2.02 $\pm$ 0.22 & 1.81 $\pm$ 0.14 & 4.00 $\pm$ 0.20 & 1.05 \\ 
\hline
XTE J1739-302 & 39 & 1.24 $\pm$ 0.19 & 1.25 $\pm$ 0.14 & 3.02 $\pm$ 0.25 & 2.83 $\pm$ 0.10 & 1.11 \\ 
\hline \hline
\end{tabular}
\end{table*}

\newpage

\begin{table*}[h]
\caption{Noise model parameters - BeXBs.}
\label{tab3:sectorModelParamsBeXB}
\centering
\begin{tabular}{c|c|c|c|c|c|c}
Source & Sector & $\alpha_0 / 10^{-3}$ & $f_c$ [\cd] & $\gamma$ &  $\alpha_w / 10^{-5}$ & $\chi^2_{red}$ \\
 \hline \hline
RX J0146.9+6121 & 58 & 0.52 $\pm$ 0.09 & 1.35 $\pm$ 0.28 & 1.80 $\pm$ 0.18 & 1.46 $\pm$ 0.17 & 1.10 \\ 
\hline
X Per & 18 & 0.28 $\pm$ 0.03 & 3.66 $\pm$ 0.35 & 2.55 $\pm$ 0.23 & 0.84 $\pm$ 0.12 & 0.98 \\
X Per & 43 & 0.46 $\pm$ 0.06 & 2.59 $\pm$ 0.33 & 2.16 $\pm$ 0.20 & 1.11 $\pm$ 0.17 & 0.94 \\
X Per & 44 & 0.64 $\pm$ 0.08 & 2.32 $\pm$ 0.29 & 2.17 $\pm$ 0.17 & 0.81 $\pm$ 0.15 & 0.83 \\
X Per & 70 & 0.31 $\pm$ 0.06 & 1.54 $\pm$ 0.40 & 1.45 $\pm$ 0.17 & 0.34 $\pm$ 0.20 & 1.08 \\
X Per & 71 & 0.28 $\pm$ 0.04 & 2.41 $\pm$ 0.37 & 1.81 $\pm$ 0.15 & 0.00 $\pm$ 0.11 & 1.26 \\
X Per & Mean & 0.42 $\pm$ 0.02 & 2.73 $\pm$ 0.12 & 2.14 $\pm$ 0.07 & 0.72 $\pm$ 0.06 & 1.50 \\ 
\hline
4U 1036-56 & 10 & 0.99 $\pm$ 0.17 & 1.35 $\pm$ 0.28 & 1.70 $\pm$ 0.15 & 1.40 $\pm$ 0.28 & 1.07 \\
4U 1036-56 & 36 & 0.86 $\pm$ 0.12 & 1.89 $\pm$ 0.29 & 2.04 $\pm$ 0.19 & 2.39 $\pm$ 0.27 & 0.94 \\
4U 1036-56 & 37 & 1.29 $\pm$ 0.22 & 1.32 $\pm$ 0.25 & 1.83 $\pm$ 0.16 & 2.05 $\pm$ 0.28 & 0.99 \\
4U 1036-56 & 63 & 1.11 $\pm$ 0.17 & 1.50 $\pm$ 0.24 & 1.95 $\pm$ 0.15 & 1.22 $\pm$ 0.19 & 1.13 \\
4U 1036-56 & 64 & 1.25 $\pm$ 0.22 & 1.21 $\pm$ 0.24 & 1.78 $\pm$ 0.14 & 1.38 $\pm$ 0.24 & 1.08 \\
4U 1036-56 & Mean & 1.25 $\pm$ 0.08 & 1.44 $\pm$ 0.10 & 1.85 $\pm$ 0.06 & 1.94 $\pm$ 0.11 & 1.19 \\ 
\hline
4U 2206+543 & 16 & 1.04 $\pm$ 0.15 & 1.76 $\pm$ 0.28 & 1.85 $\pm$ 0.14 & 0.71 $\pm$ 0.25 & 0.96 \\
4U 2206+543 & 17 & 1.03 $\pm$ 0.15 & 1.89 $\pm$ 0.27 & 2.08 $\pm$ 0.18 & 2.15 $\pm$ 0.26 & 0.84 \\
4U 2206+543 & 56 & 0.75 $\pm$ 0.09 & 2.51 $\pm$ 0.31 & 2.09 $\pm$ 0.15 & 0.44 $\pm$ 0.20 & 1.07 \\
4U 2206+543 & 57 & 0.91 $\pm$ 0.13 & 1.95 $\pm$ 0.30 & 1.84 $\pm$ 0.14 & 0.30 $\pm$ 0.24 & 1.04 \\
4U 2206+543 & Mean & 1.06 $\pm$ 0.06 & 1.97 $\pm$ 0.12 & 1.94 $\pm$ 0.06 & 0.99 $\pm$ 0.11 & 1.15 \\ 
\hline \hline
\end{tabular}
\end{table*}

\begin{table*}[h]
\caption{Noise model parameters - Others.}
\label{tab3:sectorModelParamOthers}
\centering
\begin{tabular}{c|c|c|c|c|c|c}
Source & Sector & $\alpha_0 / 10^{-3}$ & $f_c$ [\cd] & $\gamma$ &  $\alpha_w / 10^{-5}$ & $\chi^2_{red}$ \\
\hline  \hline
XTE J0421+560 & 19 & 3.25 $\pm$ 0.94 & 0.41 $\pm$ 0.12 & 1.68 $\pm$ 0.12 & 2.27 $\pm$ 0.19 & 0.99 \\
XTE J0421+560 & 59 & 3.40 $\pm$ 1.07 & 0.34 $\pm$ 0.11 & 1.61 $\pm$ 0.11 & 1.85 $\pm$ 0.17 & 0.91 \\
XTE J0421+560 & 73 & 2.65 $\pm$ 0.80 & 0.42 $\pm$ 0.14 & 1.46 $\pm$ 0.12 & 1.72 $\pm$ 0.27 & 0.90 \\
XTE J0421+560 & Mean & 3.43 $\pm$ 0.51 & 0.39 $\pm$ 0.06 & 1.57 $\pm$ 0.06 & 2.19 $\pm$ 0.11 & 1.09 \\
\hline
4U 1954+319 & 14 & 0.11 $\pm$ 0.03 & 1.00 $\pm$ 0.41 & 1.50 $\pm$ 0.33 & 1.61 $\pm$ 0.13 & 1.06 \\
4U 1954+319 & 41 & 0.17 $\pm$ 0.03 & 5.56 $\pm$ 1.04 & 1.35 $\pm$ 0.37 & 0.00 $\pm$ 1.23 & 1.07 \\
4U 1954+319 & 54 & 0.06 $\pm$ 0.00 & 6.31 $\pm$ 0.33 & 6.00 $\pm$ 1.10 & 1.47 $\pm$ 0.06 & 1.04 \\
4U 1954+319 & 55 & 0.09 $\pm$ 0.02 & 1.11 $\pm$ 0.34 & 1.62 $\pm$ 0.26 & 0.77 $\pm$ 0.07 & 1.09 \\
4U 1954+319 & 74 & 0.13 $\pm$ 0.04 & 0.82 $\pm$ 0.54 & 0.97 $\pm$ 0.20 & 0.17 $\pm$ 0.22 & 1.06 \\
4U 1954+319 & 75 & 0.07 $\pm$ 0.02 & 2.34 $\pm$ 0.82 & 1.20 $\pm$ 0.27 & 0.23 $\pm$ 0.21 & 1.01 \\
4U 1954+319 & Mean & 0.14 $\pm$ 0.01 & 1.66 $\pm$ 0.30 & 0.99 $\pm$ 0.09 & 0.43 $\pm$ 0.18 & 1.79 \\ 
\hline \hline
\end{tabular}
\end{table*}

\end{appendix}

\end{document}